\documentclass[10pt,a4paper,twoside]{article}
\usepackage{epsfig}
\usepackage{baltlat6}
\long\def\symbolfootnote[#1]#2{\begingroup%
\def\thefootnote{\fnsymbol{footnote}}\footnote[#1]{#2}\endgroup}

\begin{document}
\ \
\vspace{0.5mm}
\setcounter{page}{1}
\vspace{8mm}

\titlehead{Baltic Astronomy, vol.\,xx, xxx--xxx, 2011}

\titleb{TOWARDS THE AUTOMATIC ESTIMATION OF GRAVITATIONAL LENSES' TIME DELAYS}

\begin{authorl}
\authorb{A.~Hirv}{},
\authorb{N.~Olspert}{} and
\authorb{J.~Pelt}{}
\end{authorl}

\begin{addressl}
\addressb{}{Tartu Observatory, T\~{o}ravere, 61602, Estonia}
\end{addressl}

\submitb{Received: 2011 April 26; accepted: 2011 May 23}

\begin{summary} Estimation of time delays from a noisy and gapped data is 
one of the simplest data analysis problems in astronomy by its formulation. But as history of 
real experiments show, the work with observed data sets can be quite complex and evolved. 
By analysing in detail previous attempts to build delay estimation algorithms we try to 
develop an automatic and robust procedure to perform the task. To evaluate and compare different 
variants of the algorithms we use real observed data sets which have been objects of past 
controversies. In this way we hope to select the methods and procedures which have highest 
probability to succeed in complex situations. As a result of our investigations we propose an 
estimation procedure which can be used as a method of choice in large photometric experiments. 
We can not claim that proposed methodology works with any reasonably well sampled input data 
set. But we hope that the steps taken are in correct direction and developed software is truly 
useful for practising astronomers. \end{summary}


\resthead{Towards the automatic estimation of time delays}
{A.~Hirv, N.~Olspert, J.~Pelt}

\sectionb{1}{INTRODUCTION}

There are many astrophysical applications where number of time dependent values are measured 
and we suspect that observed time series are (may be distorted) replicas of the one and the 
same source curve. The time delays in gravitational lens observations and reverberation lags in 
high energy astrophysics are not the only examples, but probably the most commonly occurring. 
The problem of time delay estimation from noisy and gapped data is therefore one of the 
most important data processing and analysing tasks for observing astronomer.

In this paper we formulate the delay estimation problem, describe shortly the known methods 
used to estimate time delays, and propose some refinements to the old known methods. Our basic 
goal here is to connect the good sides of different methods with our improvements and to move
towards {\it automatic} estimation of the time delays. The automaticity 
we are talking about is considered along two lines: un-attended data analysis for large 
photometric experiments (see for example Oguri \& Marshall 2010), synoptic surveys (Coe \& Moustakas 2009)  
or meta-analyses (Paraficz \& Hjorth 2010) and sure-kill approach for general practising astronomer. We are well 
aware, that fully automatic approach is still a nice dream of things to come, nevertheless we 
believe that detailed implementation of proposed here methods brings us closer to final target.

When comparing different methods and evaluating their performance we will use somewhat 
different from usual approach. Instead of using artificially built data models and Monte-Carlo 
style massive calculations we apply new modifications to actually measured data sets, 
especially to those ones for which previous authors obtained differing results. In doing so we 
hope to concentrate our attention to the real life situations. It is very often so that 
theoretically sound method can give wrong answers due to the minor, but hidden, peculiarities 
in observed data. Automatic methods must be robust against such errors or warn users about 
possibility of miscalculation.

The paper is organised as follows. We begin with formulation of our problem in mathematical 
terms and describe caveats involved. Then the short overview of the well known methods follows. 
Our panorama of the methods is representative, certainly not exhaustive. Then we present some 
important modifications to the methods used so far and describe their performance using different
sets of observed data. We try to single out a more or less general and 
robust approach to time delay estimation problem to be used as a recipe for field-workers.

\sectionb{2}{TIME DELAYS}

We are dealing here with a problem where a certain physical process, which can be described by
continuous (in time) variable $g(t)$, is observed through different channels.
Because of different flight paths the total flight times $\Phi_r,r=1,\dots,R$
differ. Consequently, we can only measure replicas of the source curve with different delays:
\begin{equation}
f_r (t) = F(g(t - \Phi_r)),{\rm{  }}r = 1,2,\dots,R.
\label{variability_to_be_observed}
\end{equation}
Additional distortions (physical and instrumental) are depicted here using the function $F$
(the exact form of it depends on particular experiment). For a fully resolved case we will
have in total $R$ continuous curves $f_r(t)$ corresponding to images $f_1,\dots,f_R$.
We can observe the values of $f_r(t)$ at certain moments of time $t_i,{\rm{ }}i=1,2,\dots,N$.

The absolute flight times $\Phi_r,r=1,\dots,R$ can not be measured directly from light curves.
When talking about time delay estimation, we can define and measure the differential time delay
$\Delta t_{o,p}=\Phi_p-\Phi_o$ between each pair of images $f_o,f_p$.
The time delay $\Delta t_{o,p}$ is positive if the variability of the image $o$ is
preceding the variability of the $p$ image.

In the present paper we concentrate our attention to the time delay estimation between two
resolved light curves. For the blended cases see for example Geiger \&
Schneider (1996), Hirv et al. (2007a) and Hirv et al. (2007b).

\sectionb{3}{ESTIMATION METHODS}

Roughly the different methods and approaches to delay estimation can be divided into three 
groups: cross-correlation based, these which use dispersion spectra and finally methods based 
on certain interpolation or approximation schemes. To unify treatment below we use the 
following notation. There are two sets of data points $A$ and $B$ (time, measured value and 
standard deviation for every point) with $N_A$ points $t_i^*,a_i^*,\sigma_i^*, i=1,2,\dots,N_A$ 
and $N_B$ points $t_j^*,b_j^*,\sigma_j^*, j=1,2,\dots,N_B$. It is not assumed that $N_A=N_B$. 
If the observer given estimates for standard deviations are missing then we can set 
$\sigma_i^*=1$ and $\sigma_j^*=1$ (in the sense of relative weights). When comparing
two data sets we often need to {\it adjust} 
the data to take into account time delay, differences in magnification, baseline levels etc. 
For adjusted data sets we will use the simplest notation: if both sets are treated separately 
then we have triples $t_i,a_i,\sigma_i$ and $t_j,b_j,\sigma_j$ and if the sets are combined 
using certain trial time delay $\tau$ we use triples 
$t_l,y_l,\sigma_l,{\rm{~~}}l=1,2,\dots,L$. In some cases $L=N_A+N_B$ but not always.

In some formulae the notation for statistical weights $W=1/\sigma^2$ (with proper indexes) is 
more appropriate.

\subsectionb{3.1}{Methods based on cross-correlation}

Two continuous curves $a(t)$ and $b(t)$ can be correlated for various delays $\tau$ by computing
\begin{equation}
CF(\tau ) = \frac{{{\bf{E}}\{ [a(t) - \bar a][b(t + \tau ) - \bar b]\} }}{{\sigma _a \sigma _b }},
\end{equation}
where application of ${\bf{E}}\{\dots\}$ denotes taking statistical expectation, $\bar a$, $\bar b$ 
are mean values (estimated or known) and $\sigma_a$ and $\sigma_b$ are corresponding standard 
deviations. It is hoped for that correct delay $\tau_{AB}$ will reveal itself as the strongest or 
at least a major maximum in the correlation curve.

There are many ways to approximate notion of correlation function for discrete time series. For 
instance we can define certain fixed step (say $t_l=l\delta t,l=0,1,\dots,L-1$) grid in time 
and map every observed point to the nearest grid point. Using now standard definition for a 
discrete correlation function we can compute certain approximation. There is no need to say 
that for our sparse data sets the resulting correlation function estimate will probably have 
gaps and considerable scatter due to the fact that the large part of the pairs to be correlated 
is missing. Sometimes it is proposed that we can add to our data sets artificial points which 
are obtained by linear interpolation, see for instance Gaskell \& Sparke (1986) or 
Gaskell \& Peterson (1987). This approach, even when useful in some contexts, can 
not be used for a data with significant gaps.

More common and often used is an approach proposed in Edelson \& Krolik (1988). 
First, for each pair of observations they define:
\begin{equation}
F_{ij}  = \frac{{(a_i  - \bar a)(b_j  - \bar b)}}{{\sqrt {(\sigma _a^2  - e_{a_i}^2 )(\sigma _b^2  - e_{b_j}^2 )} }},
\end{equation}
where $e_{a_i}^2$ and $e_{b_j}^2$ are measurement errors for $A$ and $B$ set correspondingly. Then the 
moving window with width $\Delta t$ is used to compute the so called {\it Discrete 
Correlation Function} (DCF):
\begin{equation}
DCF(\tau ) = \frac{{\sum\limits_{i,j} {S_{ij} F_{ij} } }}{{\sum\limits_{i,j} {S_{ij} } }},
\end{equation}
with inclusion condition:
\begin{equation}
S_{ij}  = \cases{ 
   1,& {\rm{  when }}$|t_i  - t_j  - \tau | \le \Delta t /2$, \cr
   0,& {\rm{  otherwise.}}  \cr
   }
\end{equation}
Depending on circumstances we can compute $DCF(\tau)$ for overlapping windows or just for a row 
of non overlapping but fully covering set of windows. The free parameter of the procedure - 
width of the moving window $\Delta t$ - is chosen as a compromise value to get enough 
resolution when trading it against statistical stability. Without certain objective method to 
fix it we can not consider $DCF$ computation as an automatic procedure.
   
In the original formulation of the $DCF(\tau)$ the means and dispersions for computing $F_{ij}$ 
values are global, they are computed for the full data sets $A$ and $B$. We can also consider a 
form of the correlation function where these values are computed separately - for the each bin 
(see Leh\'ar et al. 1992; Gil-Merino et al. 2002). This allows to take into 
account possible nonstationarity of the underlying processes. However, the problem with freely 
chosen bin size remains.

The peaks which show up in different forms of $DCF(\tau)$-s tend to be shallow and noisy. In 
Goicoechea et al. (1998) the authors propose to match shifted in $\tau$ 
autocorrelation functions with cross-correlation curves. In this case the peak position will be 
estimated using the data from the full correlation curves. This approach works only for the 
cases where structure of the curves under discussion is unimodal or nearly so.

In modification proposed by Alexander (1997) the moving window length is not fixed 
in time, but is determined by fixed number of observation pairs in the bin. This peculiarity 
allows to formulate resulting statistics in the format amenable to precise statistical 
characterisation. Unfortunately, this scheme can heavily fail when data sets contain long gaps.

All $DCF$ based methods do not take explicitly into account neither possibility of microlensing 
nor effects from long data gaps. However, they do not involve any interpolation or 
extrapolation and consequently the probability to get totally wrong results because of 
``fitting data into gaps'' is quite low. For some schemes of correlation analysis it is 
possible to use Fast Fourier Transform based methods to speed up computations (see 
Moles et al. 1986 and Scargle 1989).

\subsectionb{3.2}{Dispersion spectra}

In estimating the time delay we may adjust our
data set according to fit parameters -- trial time delay $\tau$, different magnification $m$ and
baseline shift $h$. So we can use one of the input series in adjusted form $b_j=mb_j^*+h$.
We can compute the simplest dispersion spectrum quite similarly to $DCF$. We just
define differences:
\begin{equation}
D_{ij}(m,h)=\frac{(a_i-b_j)^2}{2},
\end{equation}
and form $\tau$ dependent function
\begin{equation}
DS(\tau,m,h)=\frac{{\sum\limits_{i,j} {S_{ij} D_{ij} } }}{{\sum\limits_{i,j} {S_{ij} } }},
\end{equation}
where the inclusion condition $S_{ij}$ is the same as above.

The dispersion spectrum $DS(\tau,m,h)$ depends on $m$ and $h$. As we are interested in time delay,
the final dispersion spectrum can be computed by performing minimisation:
\begin{equation}
DS(\tau ) = \mathop {\min }\limits_{m,h} DS(\tau ,m,h).
\end{equation}
Methods using dispersion spectra allow also easily to take into account unequal quality of 
different observations (by introducing weights into squared differences). The full range of different 
implementations of dispersion spectra is presented in Pelt et al. (1996).

The averaging bin size $\Delta t$ is still a free parameter for $DS$-type methods and 
consequently we are not too much better off if to compare with $DCF$-style methods.

\subsectionb{3.3}{Interpolation based methods}

To fill gaps in observed data series we can use different interpolation or approximation 
schemes. In principle it is possible to fit certain model curves (polynomials, splines etc) to 
the both curve and then compare continuous (or regularly sampled) model curves. However, most 
often the researchers use methods where model for source curve is built and both observed 
sequences are then fitted (with proper delay or delays) to the model curve (see for instance 
Leh\'ar et al. 1992; Barkana 1997; Burud et al. 2001, Cuevas-Tello et al. 2006). In these methods 
statistical weights can be easily used.

The resolution -- statistical stability trade-off for interpolation methods is achieved by 
proper (but basically, not automatic) choice of the model form (polynomial degree, smoothing kernel, number of 
nodes for splines etc). The microlensing effects or other low-frequency disturbances can not 
be detected easily, because they will be hidden in the common model curve for both light 
curves. Consequently, in some contexts it would be useful to interpolate (or approximate) input 
curves separately. Then the misfit between them will indicate possible distortions.

Most important deficiency of the interpolation type methods is of course tendency to obtain 
results with good characteristics of the fit but still indicating wrong time delays. This 
happens if data contains nearly periodic gaps. It can fairly well happen (and it indeed does so 
often) that with certain delay the observed points of the $A$ curve fit into time gaps of the 
$B$ curve.

\subsectionb{3.4}{Using optimal prediction}

Somewhat apart from other methods stands a method developed by Press et al. (1992, 
below PRH), refined by Rybicki \& Kleyna (1994) and lucidly presented by Haarsma et 
al. (1997). In principle, this method is nearest to that which can be called 
automatic. Because our discussion below heavily uses notions and ideas promoted in these papers 
we will describe the method in some detail.

We start from a model of the observed data
\begin{equation}
y(t)=s(t)+n(t),
\end{equation}
where $s(t)$ is original (source) signal and $n(t)$ is observational noise.
The observed sample vector is then
\begin{equation}
y_i  \equiv y(t_i ),{\rm{~~~}}i = 1,2,\dots,N.
\end{equation}
Here $y_i$ denotes any time series, not explicitly combined ones.
Our goal for a moment is to ``predict'' signal value for a particular time point so that 
predicted value is as close as possible to the real one or so that prediction error:
\begin{equation}
{\bf{E}}\{ e^2 (t)\}  \equiv {\bf{E}}\{ [\hat s(t) - s(t)]^2 \}
\end{equation}
is minimised. Estimate $\hat s(t)$ that is linear in the data points $y_i$:
\begin{equation}
\hat s(t) = \sum\limits_{i = 1}^N {y_i q_i } (t),
\label{base1}
\end{equation}
can be formed using $N$ ``trial'' functions $q_i(t)$. By substituting estimate
$\hat s(t)$ into expression to be minimised we can get formulae for ``trial'' functions and 
predicted value.

Using notations
\begin{equation}
\begin{array}{l}
 C_{ij}  \equiv {\bf{E}}\{ s(t_i )s(t_j )\}  \\ 
 c_i(t)  \equiv {\bf{E}}\{ s(t_i )s(t)\}  \\
 c_c(t) \equiv {\bf{E}}\{s(t)s(t) \}  \\ 
 n_i^2  \equiv {\bf{E}}\{ n(t_i )n(t_i)\}  \\ 
 B_{ij}=C_{ij}  + n_i^2 \delta _{ij}  \\ 
 \end{array}
\end{equation}
(where $\delta_{ij}=1$ when $i=j$ and $\delta_{ij}=0$ elsewhere)  
we introduce total covariance matrix ${\bf{B}}$ with elements $B_{ij}$
and two vectors ${\bf{q}}$ and ${\bf{c}}$ with elements $q_i(t)$ and $c_i(t)$. 

It occurs (see PRH), that using these notations the ``trial'' functions in the form
\begin{equation}
{\bf{q}}(t) = {\bf{B}}^{ - 1} {\bf{c}}(t),
\end{equation}
can be used to get actual ``optimal'' predictions
\begin{equation}
\hat s(t) = {\bf{y}}^T {\bf{B}}^{ - 1} {\bf{c}}(t),
\label{base2}
\end{equation}
with corresponding expected estimation errors
\begin{equation}
{\bf{E}}\{e^2(t) \}=c_c(t)-{\bf{c}}^T(t){\bf{B}}^{ - 1} {\bf{c}}(t).
\label{base3}
\end{equation}
From this point one can proceed through three different paths. 

In PRH authors show
that optimal reconstruction of $s(t)$ is equivalent to a reconstruction
that minimises the value of $\chi^2$:
\begin{equation}
\chi^2  = \sum\limits_{i,j} {y_i } A_{ij} y_j ,
\end{equation}
where the matrix ${\bf{A}}$ is the matrix inverse of the total covariance matrix
${\bf{B}}$
\begin{equation}
{\bf{A}} = {\bf{B}}^{ - 1}  \equiv \left\{ {C_{km}  + n_k^2 \delta _{km} } \right\}^{ - 1},
\end{equation}
and $y_i$-s are adjusted (shifted by time delay and magnitude difference $B$ curve combined 
with A curve, mean $\bar y$ subtracted from both) observational data. Correspondingly they use 
the following scheme to estimate the time delays:
\begin{itemize}
\item Let us assume that the underlying source process $s(t)$ is stationary and that it can be 
described by simple analytical (with small number of parameters) form.
\item For a particular trial time delay $\tau$ and other free parameters (general mean of 
the process $\bar y$, magnitude difference between two components $\Delta y_{AB}$) we can 
compute $\chi^2$ and use it as a criterion to compare different parameterisations.
\item The time delay and parameters which minimise criterion value is then used as a 
solution.      
\end{itemize}
In actual computations authors use certain optimisations and approximations to
obtain final solution. For instance they minimise $\chi^2$ along the free parameter
$\bar y$ analytically and estimate magnitude difference using ``pointwise'' fitting
of the two curves. But these are minor aspects of the adopted 
procedure.

The second method is important modification for the PRH method suggested in Rybicki \& 
Kleyna (1994) and applied by Haarsma et al. (1997). Instead of formal 
minimisation of the $\chi^2$ they take off from Gaussian model of the process with probability 
distribution of the data vector:
\begin{equation}
P({\bf{y}}) = [(2\pi )^N \left| {\bf{B}} \right|]^{ - 1/2} e^{ - 1/2\chi ^2 } .
\end{equation}
Correspondingly, for correct solution of the time delay estimation problem they propose to 
minimise {\it log likelihood} $Q$
\begin{equation}
Q = \log (\left| {\bf{B}} \right|) + \sum\limits_{i,j} {y_i } A_{ij} y_j ,
\end{equation}
where adjusted observations $y_i$ as well as determinant $\left| {\bf{B}} \right|$ depend on 
free parameters.

The third method which uses prediction ideology is somewhat naive use of the predicted
process values itself. We call this method as a ``pointwise minimisation''. 
Let us fix a particular trial time delay $\tau$.
For each component we can interpolate values at times corresponding to the observations of the 
{\it other} component and compute standard errors for
predicted values (restricting the pairing to overlapping area of the two curves). Using
obtained pairs we can now form a standard $\chi^2$ measure of goodness-of-fit:
\begin{equation}
\chi ^2  = \sum\limits_k {\frac{{(\hat a_k  - b_k )^2 }}{{\delta _k^2 }}}  + \sum\limits_l {\frac{{(a_l  - \hat b_l )^2 }}{{\delta _l^2 }}} ,
\label{points}
\end{equation}
where $\delta_k^2$ and $\delta_l^2$ are combined variances (observer given variance plus 
variance of predicted value). And again, the best combination, to be adopted as a solution, is 
set of parameters which minimises the $\chi^2$. This type of criterion function was used already
by PRH, but only in the context of estimating magnitude difference between two light curves.
We take (Eq.~\ref{points}) as a starting point for the criterion function for our combined time delay
estimation method.

\sectionb{4}{PROBLEMS AND MODIFICATIONS}

The choice of a proper criterion function to be minimised is only part, even if important part, 
of a full time delay estimation process. What follows is a treatment of different statistical 
and computational problems which occur during implementation of the general methods 
described above.

\subsectionb{4.1}{Normalisation}

If we compare curve $A$ with shifted in time by $\tau$ curve $B$ then there is only certain 
interval in time where both curves have observed values. For longer delays this overlap area 
is shorter and for shorter delays it is longer.

For methods with $\chi^2$ calculation for the combined curve the correct approach is not self-evident. 
Original authors (as far as we understand from PRH) use for evaluation of the different delays 
the same number of data points $L=N_A+N_B$. It may fairly well be that for an ideal case, where 
actual data is a realisation of a stochastic process whose properties match these of 
hypothesised model, this approach can be considered as correct. But for real data this is 
certainly not so.

For instance, when input data contains correlated errors due to the microlensing, the scatter 
for combined curve in overlap area is certainly higher if to compare with scatter in the parts 
where only one curve is observed. The length of the high scatter area depends on delay and this 
dependency will show up in $\chi^2(\tau)$ or $Q(\tau)$ curves. One possible solution is a 
computation of criteria for only overlapping subsets and using proper normalisation using 
degrees of freedom involved. Unfortunately this approach can sometimes fail - due to the long 
gaps in the data. The exclusion or inclusion of densely populated data parts (when changing 
trial delays) adds a certain amount of extra variability which is not essentially connected to 
the goodness-of-fit.
  
In methods where only pointwise differences (Eq.~\ref{points}) are involved it is quite easy to take this disparity into 
account. We do this just by dividing weighted sums of squared differences by the sums of weights
(see Eq.~\ref{meieDStau}).

\subsectionb{4.2}{Linear prediction}

The general principles of the linear prediction are well presented in PRH. However, for the completeness and also to
bring out problems encountered, we describe here some important details of the method.

\subsubsectionb{4.2.1}{Estimation of the correlation function} 

Prediction based methods involve assumption that underlying source
process which is observed through different channels is stationary and consequently with simple
correlation structure:
\begin{equation}
C_{ij}  = {\bf{E}}\left\{ {s(t_i )s(t_j )} \right\} \equiv C(t_i  - t_j ) \equiv C(T ),
\end{equation}
where covariance function $C(T)$ is to be estimated from data. For that purpose PRH introduce
first-order structure function $V(T)$ of the source process:
\begin{equation}
V(T)=\frac{1}{2}{\bf{E}}\left\{[s(t+T)-s(t)]^2 \right\},
\end{equation}
and then get
\begin{equation}
C(T)={\bf{E}}\left\{ s^2 \right\} - V(T),
\label{ctau}
\end{equation}
where ${\bf{E}}\left\{ s^2 \right\}$ is the estimated variance of the source process.
It is important to notice that in prediction procedures we need $C(T)$ values for a continuous 
range of argument and therefore we should have a certain parametric model for it.

From the observed data we can compute point estimates for the structure function
of the source process
\begin{equation}
v_{ij}  = \frac{1}{2}\left[ {(y_i  - y_j )^2  - n_i^2  - n_j^2 } \right],
T_{ij} = |t_i - t_j|,
\label{pnts}
\end{equation}
which can be binned and averaged. Finally a continuous parametrised model is fitted into the
binned curve to get a continuous approximation of the $V(T)$. In PRH the power-law 
type model for a structure function is postulated and consequently the linear model in log-log 
coordinates is used. Authors of the original method claim that overall procedure of time delay 
estimation is quite robust against small differences in linear model 
parameters estimated from data.

\subsubsectionb{4.2.2}{Structure function}

Following PRH -- to estimate  $V(T)$, we compute time-lags $T_{ij}$ and point estimates of
the structure function $v_{ij}$ for every
independent pair of data points. We sort $T_{ij}$ and $v_{ij}$ pairs by the value of $T_{ij}$,
bin and compute the bin averages ${\overline T_{ij}}$ and ${\overline v_{ij}}$. We average $P_{bin}$
points for every bin. Next we skip bins unsuitable for the model of $V(T)$ (see the discussion below) and compute 
${\overline v_{l}}=log({\overline v_{ij}}^{\frac{1}{2}})$ and ${\overline T_{l}}=log({\overline 
T_{ij}})$ and find the linear model for the logarithmic structure function ${\overline 
v_{l}}=a{\overline T_{l}}+b$. The particular value for the bin size $P_{bin}$ is in principle free parameter
of the procedure. However the final results practically do not depend on it.
The light curves we analysed were long enough to use $P_{bin}=85$ as in PRH, but for shorter data
sets, where there would be too few bins, $P_{bin}$ should be reduced.

Next we will discuss the problems we found in the PRH treatment of the structure function.
PRH use pointwise subtraction of observational noise to estimate $V(T)$, as shown in Eq.~\ref{pnts}.
The use of the linear model for the logarithmic structure function is certainly over-simplification.
As discussed by Hovatta et al. (2007) and Hughes et al. (1992), an ideal structure 
function of observational data should contain a plateau at the variance level of observational noise,
rising part, and finally, a plateau at the total variance level at long time-lags. The
structure function $V(T)$ of source process should begin from zero level at zero time-lag and have a
long time-lag plateau at the level of ${\bf{E}}\left\{ s^2 \right\}$.
We generally do not have time series with zero time-lags. However, the beginning of $V(T)$ estimated
from observational data may also lie on negative level in real cases, as
for nearby to the zero lag data points $y_i$ and $y_j$ point estimates in Eq.~\ref{pnts} tend to be 
negative (estimated observational dispersions can be quite large, if to compare with differences).
The $V(T)$ values can not be negative by definition. To avoid 
negative bin averages we need then~-- either rather large averaging bins, or we can skip the 
bins with negative means all together. In the original paper (PRH) the first averaging bin for 
the $A$ curve was skipped from computations but it was retained for the $B$ curve. The skipping 
of data points or adjustment of the bin size -- both methods involve manual nudging which is 
unacceptable for fully automatic methods.

The real structure function may sometimes have highly oscillating large time-lag end, that may even cause
negative slope for the linear fit (in log-log scale).
As discussed by Emmanoulopoulos et al. (2010), the
position of the upper turning point of a structure function depends not only on the underlying
process but also on the length of the time series. Consequently, the possible (oscillating)
large time-lag plateau may not be connected to the real underlying quasar variability and should
be excluded from the model.

As it was stated already in PRH, the results of optimal prediction are not sensitive to the exact 
parameters (slope and intercept) of the logarithmic stucture function. We implemented the simple linear fit model 
of the PRH, but added some modifications to the structure function building procedure.

First, to avoid problems around zero lag value we skipped from our fit all the bins that
were smaller than the squared mean observational noise level ${\overline n}^2$. Bins with bigger values are
not so sensitive to observational uncertainties and are supposed to provide information about the
power law type behaviour of the structure function. This is fully automatic step and can be always performed.

Second, we also implemented the skipping of the possible (oscillating) high level plateau from the 
linear fit. However, during concrete time delay estimation computations it occurred that this procedure 
is redundant for our test data and the results do not depend on its use. The robustness of 
the linear model assumption was also stated in the original PRH paper.

And finally we postulated high level constant plateau value at the level of asymptotic variance
${\bf{E}}\left\{ s^2 \right\}$.
In this way our structure functions can have two parts: rising part from the linear fit in log-log plane and
horizontal plateau at variance level.

\subsubsectionb{4.2.3}{Variance estimation}

The variance value has an important role in the algorithmic implementation of the PRH type methods. However the
precise or well founded estimate for it is seldom available. 
For instance, to estimate ${\bf{E}}\left\{ s^2 \right\}$, authors in Rybicki \& Press (1992)
suggest to take it as $10\dots 100$ 
times the data sample variance. We found the result of optimal prediction to be quite 
insensitive to this arbitrary constant, even if it was taken as large as $10^4$ times of the measured variance. 

As we do not know more about the variance of the source process than it can be guessed from observed part of a time 
series, we estimate ${\bf{E}}\left\{ s^2 \right\}$ as the value of the largest bin (${\overline v_{ij}}$) of the 
structure function. For the large sample of concrete computations this occurred good enough and correlation matrices 
were invertible. However, in the final code we also allow iterative doubling of the variance value, until to the 
point where correlation matrices can be correctly inverted. This procedure is again fully automatic and does not need 
manual nudging.

Having now the value for variance we can fix final form for a structure function.
First, we calculate the time-lag $T_{max}$,
\begin{equation}
T_{max} = \left[ \frac{{\bf{E}}\left\{ s^2 \right\}}{10^{2b}}\right] ^{\frac{1}{2a}},
\end{equation}
where the linear model of $V(T)$ in log-log coordinates reaches the $log({\bf{E}}\left\{ 
s^2\right\}^{\frac{1}{2}})$ value and turn our model to plateau of ${\bf{E}}\left\{ s^2 \right\}$
for longer time-lags. We can do that, since we do not know anything
about the structure function above the estimated ${\bf{E}}\left\{ s^2 \right\}$ level.
The final model for the structure function of the underlying process $V(T)$ is now:
\begin{equation}
V(T)  = \cases{
   10^{2b}T^{2a},& {\rm{  when }}$0 \le T \le T_{max}$, \cr
   {\bf{E}}\left\{ s^2\right\},& {\rm{  otherwise.}}  \cr
   }
\end{equation}
It is clear from the definition that our structure functions and also corresponding
correlation functions are always positive.

\subsectionb{4.3}{Fitting data into gaps}

As we saw above the different time delay estimation schemes can be divided into two 
classes.

In the first class the input data sets $A$ and $B$ are merged using trial delay, and combined 
data set is used as it is -- without taking into account the origin ($A$ or $B$) of data points. 
Different delays are then compared by modelling of the combined data set using certain 
analytical (polynomials, splines) or statistical ($\chi^2$,$Q$) models and criteria.

In the second class the original data points enter into
estimation scheme only in  pairs where one point is from $A$ curve and the other one from $B$ curve.

The principal difference of the two schemes reveals itself in the cases where input data has 
long and more or less periodic gaps in it (say, due to the skipping of certain seasons, when 
observations are not feasible). For some trial delays it can now happen that for this 
particular shift in time the observations from curve $A$ happen to fit into gaps of the $B$ 
curve. The first class methods can be quite happy with this, the general scatter around 
continuous model is at the level of the observational errors and only these parts of the 
combined curve where data points are mixed add extra scatter. As a result there is quite high 
probability to get spurious minima in the criterion curves. The susceptibility of the data to 
such distortions can be estimated by computing {\it data windows} either in the form proposed 
in PRH (Figure~8) or as a pair count spectra as this is done in Pelt et al. (1994).

The second class of methods and the particular combined method to be proposed in this paper 
overcomes the problem of gaps in an obvious way. Even the long and continuous
stretches of the {\it one} input curve do not have any effect on final dispersion
estimates. We think that this is one of the important properties of the new method.

\subsectionb{4.4}{Decorrelation length}

The long gaps in input data sets, even if not periodic, are of grave concern from another 
point of view too. In methods where only data point pairs enter into valuations, the pairs 
whose time moments differ too much can be of great influence. However, the probability that $a_i$ 
point and corresponding $b_j$ point are correlated is quite low if time difference $t_i-t_j$ is 
long enough. The term {\it decorrelation} is often used in this context. The inclusion parameter
$\Delta t$ introduced in the DCF and DS methods is used to skip all the data point 
pairs from our statistics if the time difference exceeds this prescribed value.  

Using the pointwise minimisation (Eq.~\ref{points}), we do not need either inclusion condition $S_{ij}$, nor 
inclusion parameter $\Delta t$. In the computation of the criterion function we can always use pairs of observed and 
interpolated points from different curves at the same time moments and we do not need to include pairs with longer 
distances in time. In this way we can get rid off from another free parameter.
 
\subsectionb{4.5}{Correlated errors}

In the context of gravitational lens research the time delay estimation is often complicated by 
the feature called microlensing. From the mathematical point of view this means that the two 
curves to be compared are not exactly similar but one or both of them contain extra component 
and we can not hope to achieve perfect fit of the two components, even for a correct delay. 

The possibility of extra nuisance components is generally ignored and only perfect matches are seeked
for. In this case the extra components (as supposed) are included into schemes as a part of 
observational noise. But often the complications can be severe enough to spoil whole analysis. 
This is especially true when after matching the final difference curve will have long and 
systematic excursion away from a mean (zero) level. These kinds of correlated errors indicate 
that we need to work with models where possible long term low frequency components are included 
somehow into matching scheme. For instance in Kochanek et al. (2006), Courbin et al. (2010) authors use separate polynomial models
to describe intrinsic (source) variability and extrinsic variability (microlensing). It is also possible to take into 
account low frequency trends by preprocessing input data sets. In this case the low degree polynomials are fitted into
both light curves before matching procedure (see Pelt et al. 1994).

As we do not know, which image is affected by microlensing or other distortions,
immediate subtracting of smooth trend models from observed data is not the best solution. However,
if the selected trial time delay $\tau$ is correct, we may suppose that the variability of the
{\em difference curve} of the appropriately shifted $A$ and $B$ sequences can contain smooth trend component. 

Inserting this idea and proper normalisation into Eq.~\ref{points} we get a following match criterion
(or {\it combined dispersion spectrum}) for trial time delay:
{\setlength\arraycolsep{2pt}
\begin{eqnarray}
CDS(\tau ) & = & \min_{p_1,p_2,\dots,p_P}\frac{1}{2} \Bigg[{\sum\limits_{k} {[\hat a_k-b_k-h(p_1,p_2,\dots,p_P,t_k)]^2 W_{k}}
	\over \sum\limits_{k}{W_{k}}} +{}
		\nonumber\\
	  &   & {}+ {\sum\limits_{l} {[a_l-\hat b_l-h(p_1,p_2,\dots,p_P,t_l)]^2 W_{l}} \over \sum\limits_{l}{W_{l}}}\Bigg],
\label{meieDStau}
\end{eqnarray}
}
where $h(p_1,p_2,\dots,p_P,t)$ is a smooth time dependent trend model (polynomial or spline) with parameters
$p_1,p_2,\dots,p_P$. The combined weights $W_{k}$
are calculated as
\begin{equation}
W_{k}={W_{\hat a_k} W_{b_k} \over W_{\hat a_k}+W_{b_k}},
\end{equation}
where $W_{\hat a_k}$ and $W_{b_k}$ are weights of predicted and observed points from $A$ and $B$ curve respectively.
(The combined weights $W_{l}$ are calculated in the similar way.)

Note, that as we work with data in magnitudes,
we assume the magnification $m\equiv 1.0$. Adjusting the data for the baseline shift is included in the procedure of
subtracting the polynomial trend ($P=0$ corresponds to the constant shift).
We have divided the criterion by $2$ to get it as the estimator of the dispersion.
The trend parameters $p_1,p_2,\dots,p_P$ can be estimated using standard least squares fit method.
The number of trend parameters $P$ is a free parameter of the procedure.

To include the extrinsic component elimination procedure into fully automatic algorithm we need a method to fix the
parameter $P$. 
Our experience with actual computations shows that the best way to do this is to perform analysis with a full range
of possible $P$ values. In most cases the combined dispersion spectra do not change significantly if
we change $P$ value. Only seldom the presence of a strong nuisance component demands
inclusion of a significant trend component. In our final computer code we always compute
full series of solutions with different trend models. Particularly we used simple polynomials with
the degrees $P=0\dots 10$. The final result of the time delay analysis is then formulated as a particular delay value 
with estimated error bars and a range of the trend parameter $P$ values for which alternative 
solutions remain inside the claimed interval. 

The proposed method for subtraction of the trend component is usable if the number of
time points in a time series is sufficient, the data set has long enough time coverage and we
do not seek for too large $\tau_{AB}$ for the given time coverage. Otherwise
the polynomial may
fit and reduce the variability of the quasar. The same happens, if we use too high polynomial
degree for the given data set.
The results should be taken with extra care if our method starts reporting large
and unstable time delays, when the degree of the polynomial is increased. The best way to get
an idea what is going on, is just to look at the final combined dispersion spectra (Eq.~\ref{meieDStau}) of different
polynomial degrees.
For most light curves
the trial time delay $\tau$ can be safely varied from $0.0$ to $\pm (time coverage)/2.5$.
Then we can get stable time delay for some subset of polynomials and may improve the result
compared to matching schemes, where correlated errors are not included. Note, that
in our combined time delay estimation method to be formulated in Section~5, $\tau$ can be
varied in even wider range~-- from $0.0$ to $\pm (time coverage)/2.0$, if we do not want to subtract
the microlensing distortions.

\subsectionb{4.6}{Error bars}

First, the sampling and observational accuracy of the time series should be as
good as possible to maximise the precision of the time delay estimation. The more inhomogeneous and
larger are the observational errors and the longer are the gaps in the light curves, the noisier is
the minimum of the combined dispersion spectrum and the more insecure is the result. Having longer and better
sampled time series will improve the picture of $CDS(\tau)$, but this will not always
compensate the lack of observational accuracy.

Robust confidence intervals for time delays can be obtained using bootstrap technique
(see Pelt et al. 1996). We can take the optimal prediction of the light curve as a model and
re-sample the residuals between observed and model curves to get bootstrap estimations of the time
delay. As such procedure may be very time-consuming and giving the most accurate error bars was not the
aim of our work, we used another idea by PRH instead: the interval of the trial time delay $\tau$, that
increases the $\chi^2(\tau)$ curve by $4$ units from its minimum $\chi^2(\tau_{AB})$, corresponds
to the $95$\% formal confidence interval of the time delay $\tau_{AB}$. In order to use this approach, we
have to rescale the minimum of our $CDS(\tau)$ curve to the value of $\chi^2(\tau_{AB})$ that can be
obtained from the {\it not} normalised version of Eq.~\ref{meieDStau}:
{\setlength\arraycolsep{2pt}
\begin{eqnarray}
\chi^2(\tau_{AB}) & = & \min_{p_1,p_2,\dots,p_P} \Bigg[{\sum\limits_{k} {[\hat a_k-b_k-h(p_1,p_2,\dots,p_P,t_k)]^2 W_{k}}}
		    +{}
			\nonumber\\
		  &   & {}+ {\sum\limits_{l} {[a_l-\hat b_l-h(p_1,p_2,\dots,p_P,t_l)]^2 W_{l}}}\Bigg].
\label{meie_hiiruut_tau_AB}
\end{eqnarray}
}
After rescaling, the $CDS(\tau)$ curve has the same normalisation in the proximity of the minimum point
as it is for $\chi^2(\tau)$. Parabola can be fitted into the neighbourhood of the $CDS(\tau)$ minimum
to make estimating the error bars easier.

\sectionb{5}{THE OUTLINE OF THE COMBINED METHOD}

Taking into account ideas discussed in Section~4, we formulate now the combined procedure for
the time delay estimation.
The implementation of the combined method can be done by introducing small
changes into existing code which implements PRH method.

\begin{itemize}
\item First we subtract mean values from both light 
curves and subtract mean time moment of one curve from the time points of both curves. 
This puts our data into general position so that irrelevant particularities of the time and amplitude measurements
will be ignored.
\item For each trial time delay $\tau$
\begin{itemize} 
\item We shift the $A$ light curve by a trial time delay
$\tau$ and select points in time-shifted curves that are in the same time domain. 
Important point here is that for different delays, domains where match can be performed are of
different length. This is taken into account by normalising in the criterion function $CDS(\tau)$.
\item
Next we interpolate using optimal prediction technique values for $A$ curve at the time points of the $B$ curve
and {\it vice versa}. We have now two curves that have the same number of data points and the same
sampling structure. For every time point we will have one original value and one interpolated value.
\item We fit smooth polynomial trend model into the difference curve to eliminate the possible
low frequency distortions due to microlensing and compute $CDS(\tau)$ value (see Eq.~\ref{meieDStau}).
\end{itemize}
\item The global minimum in the run of $CDS(\tau)$ is statistic used to select the best candidate for the true
delay value $\tau_{AB}$. This minimisation procedure can be repeated for different trend models.
\item For the established best delay values the error bars can be computed using bootstrap or described above simple
procedure.
\end{itemize}

\sectionb{6}{EVALUATION OF THE METHOD}

There are two different ways to evaluate algorithms for time delay estimation. First, we can 
generate artificial data sets with known time delays, and apply different time delay finding 
methods to the generated data. We can use simple random walk for generating light curves and 
sample them randomly or use sampling of some real time series. This approach was used in
Hirv et al. (2007a,b).

However, as our experience from previous studies suggests, algorithm that works best on
generated light curves, may not always perform well on 
real observational data. As we are unable to simulate all observational effects, method that 
is ``trained'' to work on generated data may give wrong results in real case. Hence we 
decided to use real observational data for evaluating time delay estimation procedures.

If the time series is well sampled, has sufficient time coverage and reasonably small 
observational errors, the results of applying all usable time delay finding methods should 
converge on the same value. The situation changes when we apply them onto time series of lower 
quality and shorter duration. The method, that works well and gives the same answer with
higher and lower quality data of the same object, should be recognised as more stable
and consistent.

\subsectionb{6.1}{Data sets and results}

In Table~1 we present the data sets used for testing the time delay finding algorithms. The reference, object, number of 
time points and duration of each light curve are given. In the same table we also present the time delays found by 
original authors, and our results as well. In the last column the range of accepted polynomial degrees is given. 
Note, that $\tau_{AB}$ is positive, if variability of the $A$ image is preceding variability of the $B$ image.
For QSO~0957+561 and HE~1104-1805 we have three separate time series. These light curves have different
time coverages, samplings and weight systems which make them truely useful for testing the stability of our method
against variable observational quality.

\begin{table}[!h]
\begin{center}
\vbox{\small\tabcolsep=4pt
\parbox[c]{124mm}{\baselineskip=10pt
{\normbf\ \ Table 1.}{\norm\
Data sets and results.}}
\begin{tabular}{l r r c c c}            \hline \hline
Object & Points & Duration & Original $\tau_{AB}$ & Our $\tau_{AB}$ & P \\
       &        & (days)   & (days)               & (days)          &   \\ \hline
QSO~0957+561$^1$ & 131 A,B & 2926 & 415$\pm$20 & (412$\dots$416)$\pm$6  & 1..9 \\
QSO~0957+561$^2$ & 1233 A,B & 6805 & 416.3$\pm$1.7 & (417$\dots$426)$\pm$2 & 0..9 \\
QSO~0957+561$^3$ & 97 A,B & 581 & 417$\pm$3 & 417$\pm$2 & 0..5 \\
HE 1104-1805$^4$ & 236 A,B & 1630 & -157$\pm$21 & (-160$\dots$-156)$\pm$8 & 0..10 \\ 
HE 1104-1805$^5$ & 245 A,B & 1763 & -161$\pm$7 & (-160$\dots$-159)$\pm$6 & 0..7  \\
HE 1104-1805$^6$ & 383 A,B & 3279 & -152.2$^{-2.8}_{+3.0}$ & (-161$\dots$-156)$\pm$6 & 0..9 \\
SDSS~J1004+4112$^7$ & 104 A,B & 259 & -40.6$\pm$1.8 & -40$\pm$2 & 4..10 \\
HE 0435-1223$^8$ & 143 A,D & 606 & -14.37$^{-0.85}_{+0.75}$ & (-16$\dots$-14)$\pm$2 & 1..10 \\ \hline
\end{tabular}
}
\end{center}
$^1$ Vanderriest et al. (1989);
$^2$ Schild (Pelt et al. 1998); 
$^3$ Kundi\'c et al. (1997);
$^4$ Wyrzykowski et al. (2003);
$^5$ Ofek \& Maoz (2003);
$^6$ Poindexter et al. (2007);
$^7$ Fohlmeister et al. (2008);
$^8$ Kochanek et al. (2006).
\end{table}

The important point about Table~1 is that all computations for it are done with one and the same algorithmic set up. 
Even the only free parameter of the structure function building procedure (the number of observation pairs in 
bin) was set to be $P_{bin}=85$ for all tests. The step of trial time delay $\tau$ was $1.0$ days. While in general
case $\tau$ can be varied in the range from $0.0$ to $\pm (time coverage)/2.5$, we must be aware that this may not
be the case if single predicted event is observed to establish the time delay value (see Kundi\'c et al. 1995,1997 for an
example). We choose to vary $\tau$ in the ranges used by original authors. Details about particular data sets follow.

\subsubsectionb{6.1.1}{QSO 0957+561}

As the first test data we used the three photometric time series of the most well known lens system QSO~0957+561,
published by Vanderriest et al. (1989), by Schild\symbolfootnote[1]{http://cfa-www.harvard.edu/\~{}rschild/fulldata2.txt}
and by Kundi\'c et al. (1997). All the three data sets were analysed previously multiple of times.

\begin{figure}
\vbox{
\centerline{\psfig{figure=fig01.eps,width=125truemm,height=70truemm,angle=0,clip=}}
\vspace{-.5mm}
\captionb{1} {The $\chi^2$ curve of the PRH method applied to the Vanderriest et al. (1989) data. Delay estimate
$\tau_{AB} = 536$ days. }
}
\vspace{5mm}
\end{figure}
\begin{figure}
\vbox{
\centerline{\psfig{figure=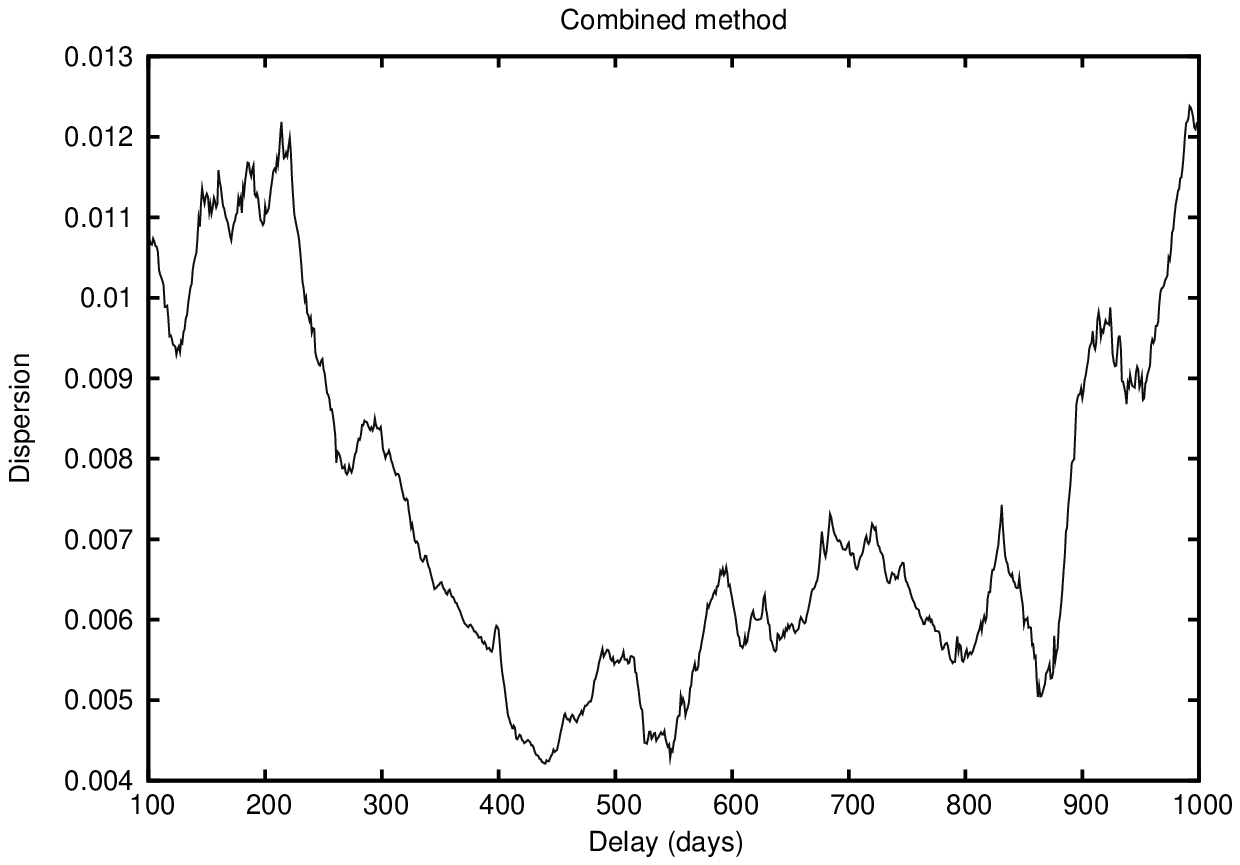,width=125truemm,height=70truemm,angle=0,clip=}}
\vspace{-.5mm}
\captionb{2} {The output curve of the combined method applied to the Vanderriest et al. (1989) data. Extrinsic variability ignored
($P = 0$), delay estimate $\tau_{AB} = 440$ days.}
}
\vspace{5mm}
\end{figure}

Vanderriest et al. (1989) got initially $415\pm 20$ days for the time delay between the $A$ and $B$ light 
curves. They used cross-correlation method on interpolated light curves and also
cross-covariance method with discrete Fourier transform to obtain that value. PRH 
analysed the same data set and obtained $536\pm 14$ days instead. 
In Pelt et al. (1994) the both delays were obtained by using different schemes of analysis.
Later Pelt et al. (1998) analysed a much longer and detailed data set provided by Schild
and reported $416.3\pm 1.7$ days for the time delay. 
In the interesting project Kundi\'c et al. (1995) found a significant drop in the photometry of the A component light curve. The observed event was used to predict time moment for a similar 
drop in the B curve. The follow up observations one and half year later confirmed the value
$417\pm 3$ days (Kundi\'c et al. 1997). The currently accepted 
time delay value for this system is still around $417$ days (see Colley et al. 2003, Shalyapin et al. 2008 and references therein). However there are another probable values around $422\dots 426$ days which are supported by some authors (Oscoz et al. 2001, Goicoechea 2002, Ovaldsen et al. 2003).

\begin{figure}
\vbox{
\centerline{\psfig{figure=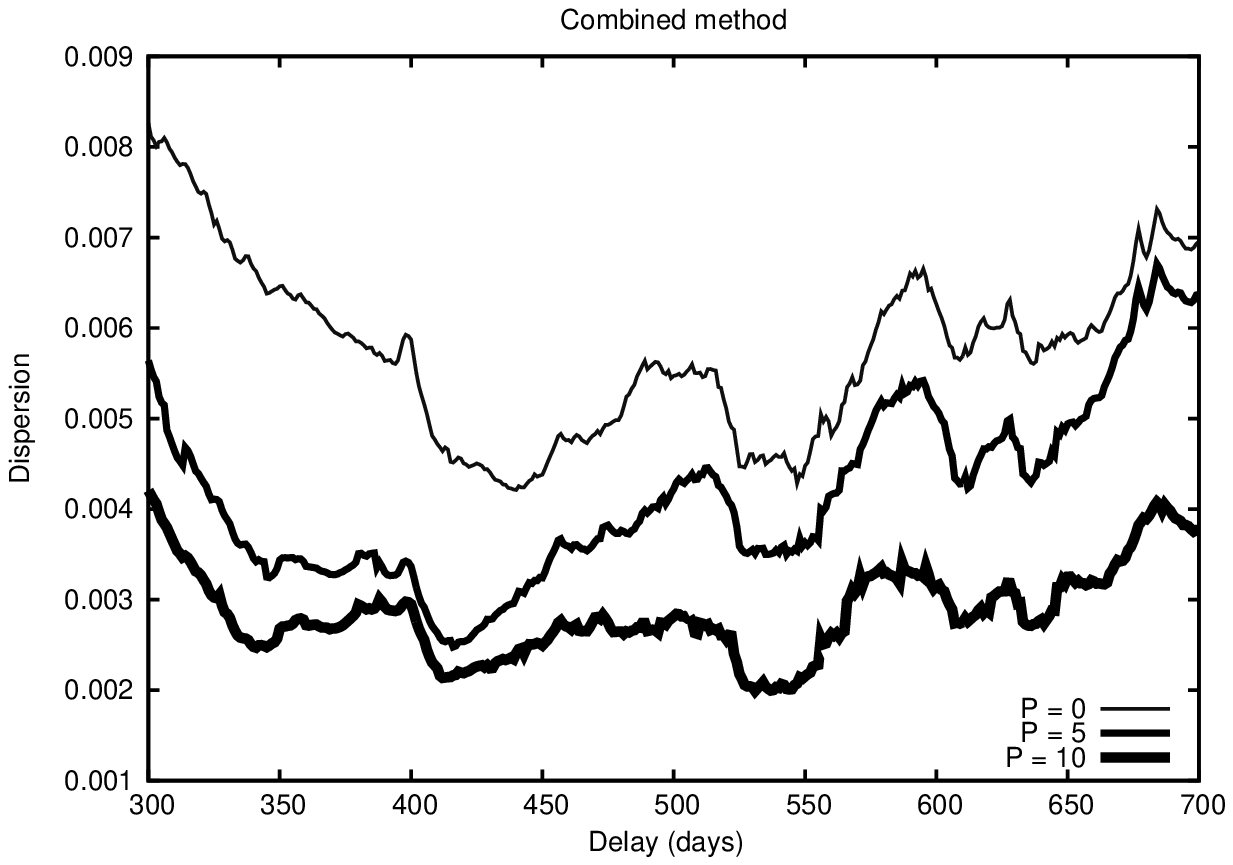,width=125truemm,height=70truemm,angle=0,clip=}}
\vspace{-.5mm}
\captionb{3} {Combined dispersion spectra for three different polynomial degrees. Vanderriest et al. (1989) data. See Table~2 for delay estimates
for different $P$ values.}
}
\vspace{5mm}
\end{figure}

Together with combined method we implemented the PRH method and applied it to the Vanderriest
et al. (1989) data. As in original paper we got $536\pm10$ days for the time delay (see Figure~1).
But, using our new approach (ignoring possible microlensing), we got $440\pm6$ days (see Figure~2).
This value is somewhat nearer to the currently accepted value but still off target. Much more clearer picture 
is revealed when we perform delay search using polynomial trend models. In Table~2 we listed our results for 
a range of polynomial degrees. The three specific spectra are also depicted in Figure~3.

\begin{table}[htb]
\begin{center}
\vbox{\small\tabcolsep=4pt
\parbox[c]{124mm}{\baselineskip=10pt
{\normbf\ \ Table 2.}{\norm\
Trend effect on time delays.}}
\begin{tabular}{c c c c c} \hline \hline
Polyn. deg. & QSO 0957+561$^1$ & HE 1104-1805$^2$ & HE 0435-1223$^3$ & SDSS~J1004+4112$^4$ \\
P	    & A,B	       & A,B 		  & A,D	             & A,B 		   \\
	    & (days)	       & (days)	          & (days)	     & (days) 	           \\ \hline
0 & 440 & -158 & -18 & -40 \\
1 & 412 & -159 & -16 & -40 \\
2 & 412 & -158 & -15 & -43 \\
3 & 415 & -160 & -15 & -38 \\
4 & 415 & -160 & -15 & -40 \\
5 & 416 & -160 & -15 & -40 \\
6 & 413 & -160 & -15 & -40 \\
7 & 416 & -160 & -15 & -40 \\
8 & 416 & -160 & -14 & -40 \\
9 & 412 & -160 & -14 & -40 \\
10 & 531 & -156 & -14 & -40 \\ \hline
\end{tabular}
}
\end{center}
$^1$ Vanderriest et al. (1989);
$^2$ Wyrzykowski et al. (2003);
$^3$ Kochanek et al. (2006);
$^4$ Fohlmeister et al. (2008).
\end{table}

From the results of our fully automatic combined method (see Table~1, Table~2 and Figures~2-3) we can conclude following:
\begin{itemize}
\item {\em Vanderriest data, no trend.} The pointwise matching using Eq.~\ref{meieDStau} gives somewhat more realistic
delay estimate if to compare with PRH method (440 against 536, true value assumed to be around 417 days). The effect of data fitting into the gaps is
not so pronounced, but result is still off target. 
\item {\em Vanderriest data, with trend model.} For degrees $P = 1\dots 9$ we got consistent set of delay values well inside of error bars of the current best estimates and also similar to the value obtained in the original paper. From what follows that original implementation of the PRH method 
was unsuccessful because of two reasons - data fitting into the gaps due to the use of global
$\chi^2$ matching criterion and also due to the leaving off possibility of microlensing. This conclusion is of separate interest (see Press \& Rybicki 1997).
\item {\em Schild's data.} Paradoxically, the most abundant and longest data series for the double quasar does not help us to fix time delay finally and sharply. We are not going to solve here this so called {\em small controversy} of the QSO 0957+561 time delay (see Goicoechea 2002, Hirv et al. 2007a, Shalyapin et al. 2008) and leave it for further studies.
\item {\em Kundi\'c's (g-filter) data.} The time delay $417$ days (also confirmed by use of the combined method) between two sharp features in the A and B light curves is used by many authors as the definitive value. However, because of {\em short controversy} we are not so convinced. Long time statistical behaviour of the light curves is quite complex and final word is not said.
\end{itemize}

\subsubsectionb{6.1.2}{Other data sets}

From Table~1 and Table~2 the results for other five data sets can be read off. One of the particular solutions is also illustrated in Figure~4.  

\begin{figure}
\vbox{
\centerline{\psfig{figure=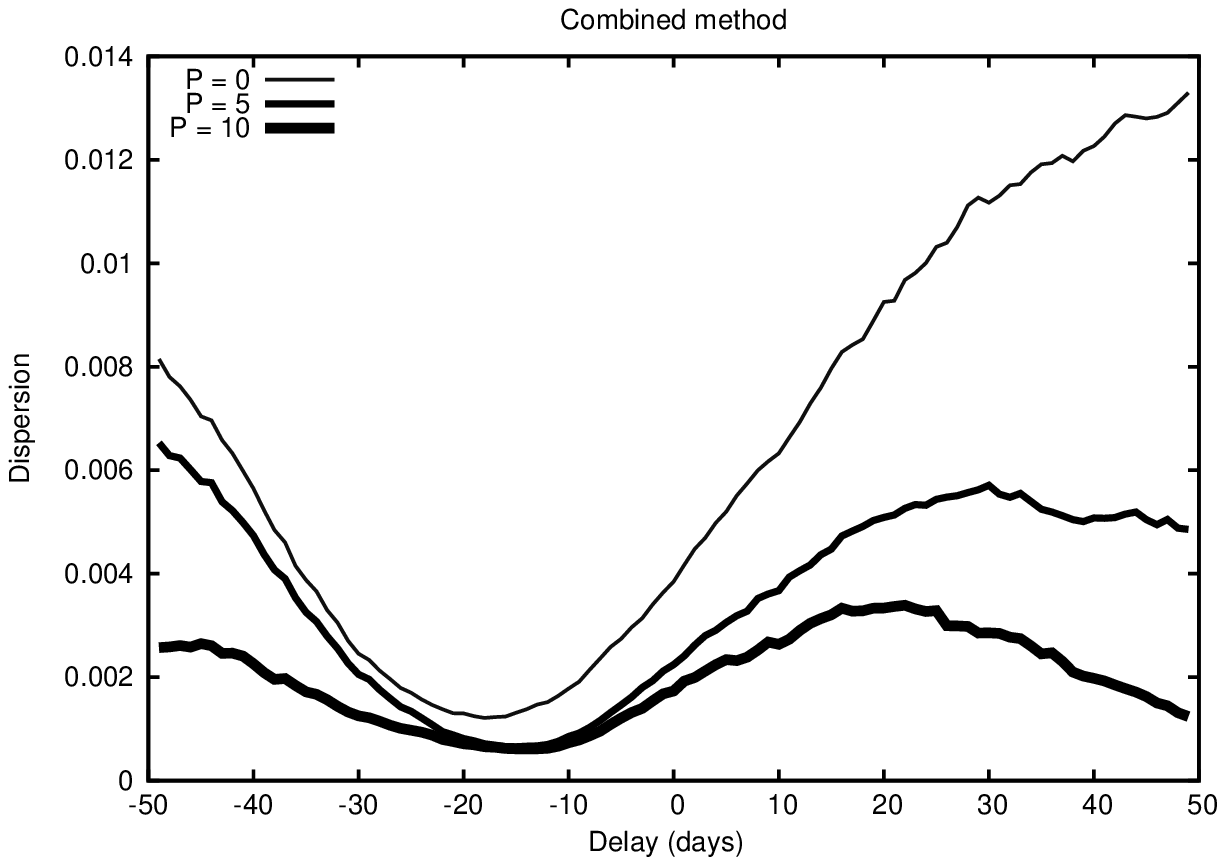,width=125truemm,height=70truemm,angle=0,clip=}}
\vspace{-.5mm}
\captionb{4} {Combined dispersion spectra for Kochanek et al. (2006) data.}
}
\vspace{5mm}
\end{figure}

We take an opportunity to stress once more again - all the results obtained are computed by
using our software as a black-box. No manual nudging, fixing certain free parameters or extra selection among different variants. Typical output of our code is just a list of delays for different trend degrees and as it is seen from Table~2 this is enough - the best estimate reveals itself as a sequence of similar or absolutely equal values in the list. Of course, there are some
important restrictions. We must have enough observations, the sampling must have a reasonably good coverage, the observational errors should not be exceedingly large etc. But these are just standard demands for a good photometry.

\sectionb{7}{DISCUSSION}

The proposed in this paper method for fully automatic time delay estimation is actually a combination of ideas from 
different previously well known methods. First we use cross-interpolation scheme which was introduced in Gaskell \& 
Sparke (1986) and Gaskell \& Peterson (1987). Then we compute actual interpolated values using linear prediction 
scheme introduced in PRH. We added to this scheme only minor improvements - rules for excluding certain bins, 
computing of variance level, as well as using the estimated variance in building the model of the structure 
function. The use of pointwise $\chi^2(\tau)$ matching criteria is ubiquitous, but not always with 
correct normalisation. And finally, the trend component fitting into the differences is implementation of ideas from 
Pelt et al. (1996). In this way the step undertaken is relatively small. However, we were somewhat amazed how 
persistently the combined method landed at or very near to the already established delay values.

\sectionb{8}{CONCLUSIONS}

As the result of our work a method for automatic time delay estimation was developed, where the number of parameters 
subjectively set by user is reduced to minimum. We have also removed many problems from methodology, which could 
lead to wrong time delay estimations (fitting data into gaps, ignoring of extrinsic variation etc). The method was 
tested to work correctly on various observed data sets. Hence, we encourage observers to use it on their own data.

\thanks{We are grateful to the referee for valuable comments. A. H. thanks Margit Hirv for great support
and inspiration. Part of this work was supported by the Estonian Science Foundation grants Nos.~6810 and 6813.}

\References

\refb Alexander T. 1997, in Astronomical Time Series, eds. D. Maoz, 
A. Sternberg \& E. M. Leibowitz, Kluwer, Dordrecht, p. 163

\refb Barkana R. 1997, ApJ, 489, 21 

\refb Burud I., Magain P., Sohy S., Hjorth J. 2001, A\&A, 380, 805

\refb Coe D., Moustakas L. A. 2009, ApJ, 706, 45

\refb Colley W. N., Schild R. E., Abajas C. et al. 2003, ApJ, 587, 71 

\refb Courbin F., Chantry V., Revaz Y. et al. 2010, arXiv:1009.1473v1[astro-ph.CO]

\refb Cuevas-Tello J. C., Ti\u{n}o P., Raychaudhury S. 2006, A\&A, 454, 695

\refb Edelson R. A., Krolik J. H. 1988, ApJ, 333, 646

\refb Emmanoulopoulos D., McHardy I. M., Uttley P. 2010, MNRAS, 404, 931

\refb Fohlmeister J., Kochanek C. S., Falco E. E., Morgan C. W., Wambsganss J. 2008, ApJ, 676, 761

\refb Gaskell C. M., Sparke L. S. 1986, ApJ, 305, 175

\refb Gaskell C. M., Peterson B. M. 1987, ApJS, 65, 1 

\refb Geiger B., Schneider P. 1996, MNRAS, 282, 530 

\refb Gil-Merino R., Wisotzki L., Wambsganss J. 2002, A\&A, 381, 428

\refb Goicoechea L. J., Mediavilla E., Oscoz A., Serra M., Buitrago J. 1998, Ap\&SS, 261, 341

\refb Goicoechea L. J. 2002, MNRAS, 334, 905

\refb Haarsma D. B., Hewitt J. N., Leh\'ar J., Burke B. F. 1997, ApJ, 479, 102

\refb Hirv A., Eenm\"ae T., Liimets T., Liivam\"agi L. J. and Pelt J. 2007a, A\&A, 464, 471

\refb Hirv A., Eenm\"ae T., Liivam\"agi L. J. and Pelt J. 2007b, Baltic Astronomy, 16, 241 

\refb Hovatta T., Tornikoski M., Lainela M. et al. 2007, A\&A, 469, 899

\refb Hughes P. A., Aller H. D., Aller M. F. 1992, ApJ, 396, 469

\refb Kochanek C. S., Morgan N. D., Falco E. E. et al. 2006, ApJ, 640, 47

\refb Kundi\'c T., Colley W. N., Gott III J. R. et al. 1995, ApJ, 445, L5

\refb Kundi\'c T., Turner E. L., Colley W. N. et al. 1997, ApJ, 482, 75 

\refb Leh\'ar J., Hewitt J. N., Roberts D. H., Burke B. F. 1992, ApJ, 384, 453

\refb Moles M., Garcia-Pelayo J. M., Masegosa J., Garrido R. 1986, AJ, 92, 1030

\refb Ofek E. O., Maoz D. 2003, ApJ, 594, 101

\refb Oguri M., Marshall P. J. 2010, MNRAS, 405, 2579

\refb Oscoz A., Alcalde D., Serra-Ricart M. et al. 2001, ApJ, 552, 81

\refb Ovaldsen J. E., Teuber J., Schild R. E., Stabell R. 2003, A\&A, 402, 891

\refb Paraficz D., Hjorth J. 2010, ApJ, 712, 1378 

\refb Pelt J., Hoff W., Kayser R., Refsdal S., Schramm T. 1994, A\&A, 286, 775 

\refb Pelt J., Kayser R., Refsdal S., Schramm T. 1996, A\&A, 305, 97 

\refb Pelt J., Schild R., Refsdal S., Stabell R. 1998, A\&A, 336, 829

\refb Poindexter S., Morgan N., Kochanek C. S., Falco E. E. 2007, ApJ, 660, 146

\refb Press W. H., Rybicki G. B., Hewitt J. N. 1992, ApJ, 385, 404

\refb Press W. H., Rybicki G. B. 1997, in Astronomical Time Series, eds. D. Maoz, 
A. Sternberg \& E. M. Leibowitz, Kluwer, Dordrecht, p. 61

\refb Rybicki G. B., Press W. H. 1992, ApJ, 398, 169

\refb Rybicki G. B., Kleyna J. T. 1994, in Reverberation Mapping of the Broad-Line Region in Active Galactic Nuclei, eds. P. M. Gondhalekar, K. Horne \& B. M. Peterson, ASP Conf. Ser., 69, 85

\refb Scargle J. D. 1989, ApJ, 343, 874  

\refb Shalyapin V. N., Goicoechea L. J., Koptelova E., Ull\'{a}n A., Gil-Merino R. 2008, A\&A, 492, 401

\refb Vanderriest C., Schneider J., Herpe G., Chevreton M., Moles M., Wlerick G. 1989, A\&A, 215, 1

\refb Wyrzykowski \L., Udalski A., Schechter P. A. et al. 2003, Acta Astronomica, 53, 229

\end{document}